# 3D Room Geometry Inference from Multichannel Room Impulse Response using Deep Neural Network

Inmo YEON[1]; Jung-Woo CHOI[2]

[1,2] School of Electrical Engineering, Korea Advanced Institute of Science and Technology (KAIST),

Republic of Korea

**ABSTRACT**

Room geometry inference (RGI) aims at estimating room shapes from measured room impulse responses (RIRs) and has received lots of attention for its importance in environment-aware audio rendering and virtual acoustic representation of a real venue. A lot of estimation models utilizing time difference of arrival (TDoA) or time of arrival (ToA) information in RIRs have been proposed. However, an estimation model should be able to handle more general features and complex relations between reflections to cope with various room shapes and uncertainties such as the unknown number of walls. In this study, we propose a deep neural network that can estimate various room shapes without prior assumptions on the shape or number of walls. The proposed model consists of three sub-networks: a feature extractor, parameter estimation, and evaluation networks, which extract key features from RIRs, estimate parameters, and evaluate the confidence of estimated parameters, respectively. The network is trained by about 40,000 RIRs simulated in rooms of different shapes using a single source and spherical microphone array and tested for rooms of unseen shapes and dimensions. The proposed algorithm achieves almost perfect accuracy in finding the true number of walls and shows negligible errors in room shapes.

Keywords: Room Geometry Inference, Room Impulse Response, Deep Neural Network

## 1. INTRODUCTION

Information on the room geometry helps solve various acoustic problems, such as spatial audio reproduction (1), sound source localization (2), and spatial audio object coding (3). Modeling accurate virtual acoustic environments and room geometry plays a crucial role in spatial audio reproduction for virtual reality and augmented reality (4, 5). A room impulse response (RIR) consists of many delayed impulses, including direct sound and early reflections, so it can be utilized to characterize the room shape.

Many RGI techniques utilizing the time of arrival (TOA) (6-9), time difference of arrival (TDOA) (10-12), and direction of arrival (DOA) (13) have been proposed. Recent studies (4, 14) successfully adopted neural networks to infer room geometry and room-acoustical parameters in shoebox-shaped rooms. Yu and Kleijn (4) demonstrated that a model trained by simulated RIRs can be adapted to real-world RIRs using transfer learning (15).

Nevertheless, most RGI techniques require prior information on the number of walls and have focused on finding the size (4, 14) or acoustical parameters (4) of walls in simple-shaped rooms such as convex shoebox rooms. Lovedee-Turner and Murphy (13) considered complex-shaped rooms such as non-convex L-shaped and T-shaped rooms; however, they used multiple RIRs measured for various source positions to satisfy the first-order reflection visibility for all walls. In this work, we demonstrate that the proposed network can infer the room geometry without prior knowledge of the number of walls, even in rooms with complex shapes where some of the first-order reflections cannot be measured by a microphone array. The DNN model is trained by RIRs between a spherical microphone array and a sound source centered at the array. The model is designed to produce two outputs involved with wall parameters and existence probability. Through the joint training using the wall parameter and existence losses, we show the possibility to infer the geometry of rooms with various shapes and numbers of walls.

---

[1] iyeon@kaist.ac.kr

[2] jwoo@kaist.ac.kr

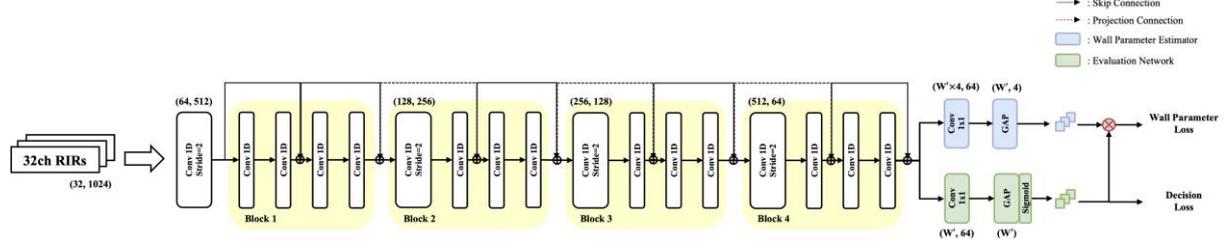

Figure 1 – Overview of the proposed network architecture ($W'$: the maximum number of inferred walls).

## 2. ROOM GEOMETRY INFERENCE

### 2.1 Problem Statement

RGI problem can be defined as finding $W$ different planes by utilizing the RIRs measured in the room enclosed by $W$ walls. Each wall constituting a room can be represented as a set $\mathcal{A}$ of points $\mathbf{r} = [x, y, z, 1]^T$ described in homogeneous coordinates and satisfying the plane equation given by:

$$\mathcal{A} = \{\mathbf{r} \in \mathbb{R}^4 \mid \mathbf{r}^T \mathbf{a} = 0\} \qquad (1)$$

where the vector $\mathbf{a} = [a_1, a_2, a_3, a_4]^T$ expresses coefficients of the plane equation. Accordingly, the goal of RGI is to find $W$ wall parameter vectors $\mathbf{a}_W$ for true walls ($w = (1, \cdots, W)$).

### 2.2 Deep Neural Networks

The proposed network consists of three sub-networks to infer room geometry (Figure 1): a feature extractor, a wall parameter estimator (WPE), and an evaluation network. The feature extractor based on ResNet (16) is responsible for extracting the features related to the room geometry from multichannel RIRs. The output of the feature extractor is fed into the wall parameter estimator and evaluation network. The WPE is a combination of 1×1 convolution and global average pooling (GAP) layers and estimates a wall parameter matrix $\tilde{\mathbf{A}} = [\tilde{\mathbf{a}}_1, \cdots, \tilde{\mathbf{a}}_{W'}]^T \in \mathbb{R}^{W' \times 4}$ for $W'$ wall candidates. The WPE strives to minimize the difference between $\tilde{\mathbf{A}}$ and the ground truth (GT) wall parameter matrix $\mathbf{A} = [\mathbf{a}_1, \cdots, \mathbf{a}_W, \mathbf{0}_{4 \times (W'-W)}]^T$. Here, the GT parameters for non-existing walls are set to zero, so the model is trained to produce zero outputs for such walls. However, even when the WPE is well-trained, there can be residual values in $\tilde{\mathbf{A}}$ for non-existing walls. To accelerate the convergence and promote parameter suppression for non-existing walls, the evaluation network predicting a wall presence probability is introduced.

The evaluation network takes the same input and has a similar layer structure as the WPE but generates only $W'$ wall existence probabilities $\hat{\mathbf{p}} = [\hat{p}_1, \cdots, \hat{p}_{W'}]^T \in \mathbb{R}^{W'}$ whose values are limited within $[0, 1]$ by a sigmoid activation function. The final estimation of the wall parameter matrix is made by weighting the estimated wall parameters in terms of the wall existence probability: $\hat{\mathbf{A}} = \text{Diag}(\hat{\mathbf{p}})\tilde{\mathbf{A}}$.

Two loss functions are defined to train networks. First, the wall parameter loss is defined as the angular loss $\gamma = 1 - |\hat{\mathbf{b}}^T \mathbf{b}|^2 / (\|\hat{\mathbf{b}}\|^2 \|\mathbf{b}\|^2)$ between $\hat{\mathbf{b}} = \text{flatten}(\hat{\mathbf{A}})$ and $\mathbf{b} = \text{flatten}(\mathbf{A})$, where flatten(·) represents the flattening operator reshaping a matrix into a vector. The second loss is the decision loss $\beta$, which is defined as binary cross entropy of the predicted wall presence probability $\hat{\mathbf{p}}$ and the GT presence probability $\mathbf{p} = [p_1, \cdots, p_{W'}]^T$ ($p_w = 1$ for true walls, $p_w = 0$ otherwise). These two losses are summed to calculate the total loss as $\mathcal{L} = \gamma + 0.1\beta$. To circumvent the order permutation problem between the GT and estimated walls, the permutation invariant training technique (17) is adopted.

## 3. EXPERIMENTAL RESULTS

### 3.1 Dataset

To build datasets using simulated RIRs, we considered an acoustically transparent spherical microphone array with 32 omnidirectional microphones arranged on a hypothetical sphere of 0.042 m radius. A single loudspeaker was then positioned at the center of the spherical microphone array to simulate a single device with an embedded loudspeaker and microphones. As shown in Figure 2, the device was positioned at the center of rooms having four different shapes: shoebox-, pentagonal-, hexagonal-, and L-shaped rooms. The walls were all planar, and the floor and ceiling were parallel to each other but perpendicular to the side walls. The size and aspect ratio of rooms were varied by randomly populating the size of a rectangular box that tightly encloses rooms in Fig. 2. The edge lengths of a rectangular box in x-, y-, and z-directions were sampled from a uniform distribution within the range $[4, 10]$, $[4, 10]$, and $[3, 5]$ m, respectively.

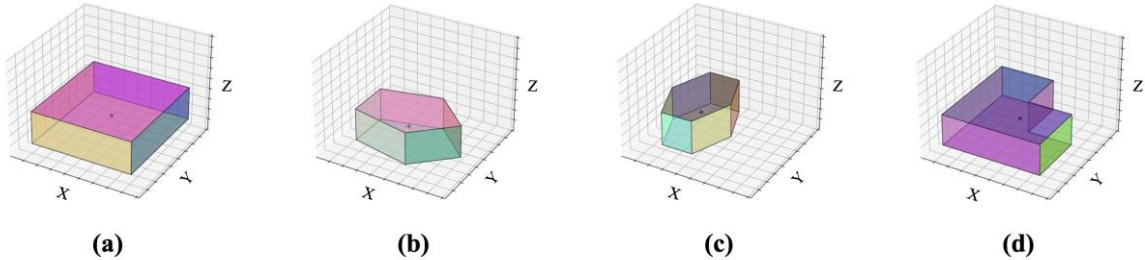

Figure 2 – Examples of simulated rooms including convex and non-convex shapes. (a), (b), (c), and (d) represent shoebox-, pentagonal-, hexagonal-, and L-shaped rooms, respectively. In particular, in (d), the first-order reflections of some walls are not visible from the microphone array.

Note that the shoebox-, pentagonal-, and hexagonal-shaped rooms are convex while the L-shaped rooms are non-convex. The first-order reflection visibility was satisfied for all walls in the convex rooms but was not fulfilled for some walls in the case of the non-convex rooms.

Multichannel RIRs were simulated by considering up to sixth-order reflections through the image source model (ISM). The length of each RIR was 1024 taps (approximately 0.1 s) at a sampling rate of 8 kHz. Finally, the dataset with 40k multichannel RIRs was constructed including 39k train data and 1k validation data. Another dataset with 500 RIRs was simulated from unseen rooms and used as a test dataset.

### 3.2 Results

Table 1 – Geometry inference performance of the proposed network

| Metric | Total | Shoebox | Pentagonal | Hexagonal | L-shaped |
|---|---|---|---|---|---|
| $ACC_W$ (%) | 100 | 100 | 100 | 100 | 100 |
| $\Delta d$ (m) | 0.01 | 0.01 | 0.01 | 0.01 | 0.01 |
| $\Delta \theta$ (degree) | 0.07 | 0.05 | 0.08 | 0.08 | 0.07 |

Three performance metrics were used to evaluate the geometry inference performance of the proposed network. The first one is the accuracy in the number of estimated walls ($ACC_W$), defined as

$$ACC_W = 100 \times \frac{1}{L} \sum_{\ell=1}^{L} \delta(\hat{\mathbf{p}}^{(\ell)}, \mathbf{p}^{(\ell)}), \text{ where } \delta(\hat{\mathbf{p}}^{(\ell)}, \mathbf{p}^{(\ell)}) = \begin{cases} 1 & \text{if } \sum_{w=1}^{W'} \text{XNOR}(\hat{p}_w^{(\ell)}, p_w^{(\ell)}) = W \\ 0 & \text{if } \sum_{w=1}^{W'} \text{XNOR}(\hat{p}_w^{(\ell)}, p_w^{(\ell)}) \neq W \end{cases} \quad (2)$$

Here, $\ell$ and $L$ denote the indices and the total number of rooms in the dataset, respectively. The second measure is the error in the shortest distance from the device to the wall ($\Delta d$) (8, 9), and the third one is the acute angle between surface normal vectors of the estimated and GT walls ($\Delta \theta$) (9, 13).

Table 1 demonstrates that the proposed network can infer the number of true walls perfectly and can estimate wall parameters with negligible errors on both metrics $\Delta d$ and $\Delta \theta$. The best performance is shown in the shoebox-shaped rooms, but the error level in the L-shaped rooms where the first-order reflections of some walls are not visible is not significantly different compared with other rooms. The proposed network shows excellent inference performance without prior assumptions for all convex and nonconvex rooms.

Despite the superior performance of the proposed method, the adaptation to various real RIR data remains a challenge. Even with techniques applied to prevent overfitting in this work, such as early stopping and GAP, overfitting could have occurred due to the same RIR simulation framework used for both the training and test dataset. In the real world, there are various factors that affect RIR, such as sound-occluding objects, diffractions and scatterings, and background noises, which were not considered in this simulation data. However, the ability to infer complex room shapes using a simple DNN model demonstrated in this work stresses that room geometry inference through the real RIRs can be realized when proper fine-tuning or noise reduction model can be implemented.

## ACKNOWLEDGEMENTS

This work was supported by National Research Council of Science and Technology (NST) funded by the Ministry of Science and ICT (MSIT) of Korea (Grant No. CRC 21011), the National Research Foundation of Korea (NRF) Grant funded by MSIT of Korea (No.1711091575), and the BK21 FOUR program through the National Research Foundation (NRF) funded by the Ministry of Education of Korea.